# Is entanglement necessary to have unconditional security in quantum bit commitment?

**Arindam Mitra**

Anushakti Abasan, Uttar Falguni -7, 1/AF, Salt Lake,

Kolkata, West Bengal, 700064, India.

Abstract: A simple un-entanglement based quantum bit commitment scheme is presented. Although commitment is unconditionally secure but concealment is not.

## Introduction

It was claimed [1] that unconditionally secure quantum bit commitment (QBC) is impossible. Recently following alternative model of quantum information processing (AMQIP) [2] we have found an entanglement-based unconditionally secure QBC scheme [3]. Here we present a similar scheme based on un-entanglement. Let us present the scheme.

Like BB-84 protocol [4] Bob transmits a sequence of $0^o$, $90^o$, $45^o$, $135^o$ polarized single photons to Alice. Alice wants to use the entire sequence to commit a single bit value. The scheme has to execute in the following order.

1. Alice measures polarization in rectilinear and diagonal bases at random. To commit a bit 0 Alice reveals results $R_i$ in the direct order. To commit bit 1 Bob reveals results in the reverse order.
2. Alice unveils bases $A_i$ in the direct order.
3. If the bit is 0 Bob will recover it from the direct order correlation in their results and if the bit is 1 he will recover it from reverse order correlation. 100% correlation can be found in the results where they measure in the same basis. To find out the direct order correlation Bob has to sort out the results where they measure in the same basis without changing the order of their results. To find out the reverse order correlation, Bob has to sort out the results after reversing the order of either Alice's results or his results.

*Security of concealment*: As soon as Alice commits her results Bob can recover 75% correlated data either in direct order or in reverse order of arrangement. From this correlation Bob can recover bit value without waiting for Alice's bases. To suppress the correlation Alice can introduce some errors in the results. Suppose after introducing 50% error in her results Alice reveals them as a mark of her commitment Due to the 50% change in Alice's results Bob will not get more than 62.5% correlation. Before unveiling step Bob can recover bit value with less confidence level.

*Security of commitment:* As Bob can recover bit value prior to unveiling step then cheating cannot be guaranteed. If Alice changes bit value after Bob's recovery of bit value Bob could detect this cheating. New bit will be different from old bit. It is easy to see [2] that noise will not help the cheating party. The commitment is logically unconditionally secure since concealment is not at all secure.

The presented scheme is not unconditionally secure quantum bit commitment-concealment (QBCC) scheme. It is an open problem to prove the possibility or impossibility of unconditionally secure QBCC without using entanglement. We conjecture that entanglement and AMQIP are necessary ingredients.